\documentstyle[12pt]{article} 
\begin{document}
\newcommand{\beq}{\begin{equation}}
\newcommand{\eeq}{\end{equation}}
\newcommand{\beqn}{\begin{eqnarray}}
\newcommand{\eeqn}{\end{eqnarray}}
\newcommand{\slp}{\raise.15ex\hbox{$/$}\kern-.57em\hbox{$\partial
$}}
\newcommand{\slA}{\raise.15ex\hbox{$/$}\kern-.57em\hbox{$A$}}
\newcommand{\slC}{\raise.15ex\hbox{$/$}\kern-.57em\hbox{$C$}}
\newcommand{\lnA}{\raise.15ex\hbox{$/$}\kern-.57em\hbox{$A_{c}^{N}$}}
\newcommand{\slB}{\raise.15ex\hbox{$/$}\kern-.57em\hbox{$B$}}
\newcommand{\bP}{\bar{\Psi}}
\newcommand{\bC}{\bar{\chi}}
 \newcommand{\hs}{\hspace*{0.6cm}}

\title{Functional integral approach to multipoint correlators in 2d critical
systems}
\author{V.I.Fern\'andez$^{a}$ and C.M.Na\'on$^{a,b}$}
\date{October 1996}
\maketitle

\def\thepage{\protect\raisebox{0ex}{\ } La Plata 96-20}
\thispagestyle{headings}
\markright{\thepage}

\begin{abstract}

\hs We extend a previously developed technique for computing spin-spin critical
correlators in the 2d Ising model, to the case of multiple correlations.
This enables us to derive Kadanoff-Ceva's formula in a simple and elegant way. 
We also exploit a doubling procedure in order to evaluate the critical  
exponent of the polarization operator in the Baxter model. Thus we provide 
a rigorous proof of the relation between different exponents, 
in the path-integral framework.
\end{abstract}
 
\vspace{3cm}
Pacs: \\ 
\hspace*{1,7 cm} 11.10.-z \\
\hspace*{1,7 cm} 11.15.-q \\
\hspace*{1,7 cm} 02.90.+p \\ 
\hspace*{1,7 cm} 05.50.+q

\noindent --------------------------------

\noindent $^a$ {\footnotesize Depto. de F\'\i sica.  Universidad
Nacional de La Plata.  CC 67, 1900 La Plata, Argentina.\\
e-mail: naon@venus.fisica.unlp.edu.ar}

\noindent $^b$ {\footnotesize Consejo Nacional de Investigaciones
Cient\'\i ficas y T\'ecnicas, Argentina.}

\newpage
\pagenumbering{arabic}

\hs Since Schultz, Mattis and Lieb \cite{SML} showed that Onsager's solution of
the two-dimensional (2d) Ising model could be simply explained in terms of a 
single Majorana fermion, there has been an increasing interest in the study of
2d statistical mechanics models by means of field-theoretical methods. In the
same vein, Luther and Peschel \cite{LP} proved that the scaling regime of the
eight-vertex (Baxter \cite{Bax}) model can be described in the continuum limit
in terms of a Thirring \cite{Thirring} Lagrangian. In this way, the 2d Ising 
and Baxter models became fruitful testing grounds for new ideas  
and computational methods.\\ 
In a previous work it has been shown how to evaluate 2-point correlators 
in 2d systems \cite{Na}, through a path-integral approach to bosonisation 
\cite{Ga}. In particular, the critical behavior of the Ising (on-line) spin-spin 
correlation function was obtained, by using a slightly modified version of the 
identity derived by Zuber and Itzykson \cite{ZI}:

\beq
{F_{2}}^{2}(x_{1},x_{2}) =<\sigma(x_{1})\sigma(x_{2})>^{2} = 
<exp\;{ \pi \int_{x_{1}}^{x_{2}}dz J_{0}(z)}> 
\label{1}
\eeq

\noindent where $J_{\mu}$ is the Dirac fermion current which is obtained out of
the original Majorana fields after squaring the correlator. $< >$ means vacuum
expectation value (v.e.v.) in a model of free massless fermion fields.\\

\hs The purpose of this note is twofold. On the one hand we extend the above
mentioned method to compute the 2n-point correlator. Thus, we provide an 
alternative derivation of Kadanoff-Ceva's formula \cite{KC} that could be
useful when considering certain non-trivial extensions of the Ising model
such as the off-critical \cite{Dotsenko} and the defected \cite{defected}
cases. On the other hand we adapt the doubling technique \cite{Ferrell} which 
led to (1), in order to calculate the correlation function of the polarization 
operator in the Baxter model \cite{Bax}. This, in turn, allows us to provide
a path-integral confirmation of the relations between different critical
exponents (those corresponding to energy-density, crossover and polarization),
a result previously established by Drugowich de Felicio and Koberle \cite{DK}
in the operator framework.\\ 

For the sake of clarity we shall begin by briefly summarizing the main points of
the spin-spin correlator calculation.
In ref.\cite{Na} the line integral in (1) was written as 

\[
 \int_{x_{1}}^{x_{2}} dz J_{0}(z) = \int d^{2}x \bP~ \slA \Psi 
\]

\noindent where $A_{\mu}$ is an auxiliary vector field with components:
\[
A_{0}(z_{0},z_{1}) = \delta(z_{0}) \theta(z_{1}-x_{1}) \theta(x_{2}-z_{1})
\]
\[
A_{1}(z_{0},z_{1}) = 0 .
\]

\noindent This simple manipulation enabled us to express the squared spin-spin 
correlator in terms of fermionic determinants:

\beq
{F_{2}}^{2}(x_{1},x_{2}) = \frac{det(i\slp + \pi\slA)}{deti\slp} \label{5}                                                                          
\eeq
\noindent where the coordinate dependence in the right hand side (r.h.s.) of 
(\ref{5}) is, of course, contained in $\slA$.\\
 Finally, one performs a change of path-integral fermionic variables 
which is chosen so as to decouple fermions from the background field $A_{\mu}$.
It is interesting to note that, in this formulation, the desired 2-point 
function is just the square root of the Fujikawa Jacobian $J_{F}$ \cite{Fu} 
associated with the transformation in the fermionic measure: 

\beq
{F_{2}(x_{1},x_{2})} ={J_{F}(x_{1},x_{2})}^{\frac{1}{2}} 
\label{6}                                                                          
\eeq
As shown in \cite{Nao}, this Jacobian must be computed with a gauge-invariant
regularization prescription in order to avoid a linear divergence (this gauge
invariance is a consequence of a symmetry in the original lattice system 
\cite{FS}). This procedure then leads to the well-known power law decay of the 
spin-spin on-line function, with exponent equal to $\frac{1}{4}$.\\

Let us now show how to extend the above depicted technique to the computation of 
the 2n-point spin correlation function at criticality. To this end, we follow 
ref.\cite{ST} where it was shown that, after squaring the correlator, each pair
of consecutive spin variables can be identified with an exponential similar to
the one appearing in (\ref{1}) (See also ref.\cite{Bo} for a very interesting 
study on the doubling procedure and the operator content of fermion fields in 
the Ising model). We can then express the squared $2n$-point correlator as

\beq
{F_{2n}}^2(x_{1},...,x_{2n})=<\prod_{i=1}^{2n} \sigma (x_{i})>^{2} =    
<\prod_{i=1,odd} exp{\; \pi \int_{x_{i}}^{x_{i+1}}dzJ_{0}(z)}> \label{7}
\eeq

\noindent where, as before, $< >$ in the r.h.s. means v.e.v. 
to be evaluated in a model of massless Dirac fermions. It is apparent 
that each line integral in (\ref{7}) can be cast in the form

\[
\int_{x_{i}}^{x_{i+1}}dzJ_{0}(z) =\int d^{2}z J_{\mu}(z)
A_{\mu}(z;x_{i},x_{i+1}) 
\]
where we have introduced the $n$ classical singular potentials

\[
A_{0}(z;x_{i},x_{i+1}) = \delta(z_{0}) \theta(z_{1}-x_{i}) \theta(x_{i+1}-z_{1})
\]

\[
A_{1}(z;x_{i},x_{i+1}) = 0 .
\]
In order to rewrite (\ref{7}) in a more compact way we construct a new vector
field $C_{\mu}$ as a simple superposition of $A_{\mu}$'s:

\beq
C_{0}(z) = \sum_{i=1, odd}^{2n-1} A_{0}(z;x_{i},x_{i+1})
\label{11}
\eeq

\beq
C_{1}(z) = 0 .\label{12}
\eeq
Thus, the $2n$-point function can be expressed in terms of fermionic 
determinants:
\beq
{F_{2n}}^{2} = \frac{det(i\slp + \pi\slC)}{deti\slp}, \label{13}                                                                          
\eeq
exactly as it happens in the $n=1$ case (See (\ref{5})), but with $A_{\mu}$ 
replaced by $C_{\mu}$.\\
The next step is to write $C_{\mu}$ in terms of scalar functions $\Phi_{c}$ 
and $\eta_{c}$ as\\
\beq
C_{\mu} = \epsilon_{\mu\nu}\partial_{\nu}\Phi_{c} + \partial_{\mu}\eta_{c}. 
\label{14}                                                                          
\eeq
Now we perform a decoupling change of path-integral fermionic variables with
chiral and gauge parameters $\Phi_{c}$ and $\eta_{c}$, respectively:\\
\beq
\Psi = e^{-\pi(\gamma_{5}\Phi_{c}+i\eta_{c})}\; \chi
\label{15}
\eeq
\beq
\bP = {\bC}\; e^{-\pi(\gamma_{5}\Phi_{c}-i\eta_{c})}
\label{16}
\eeq
A detailed computation of the Fujikawa Jacobian $J_{F}$ associated to this 
change has been given many times in the literature (See, for instance, 
ref.\cite{Ga}); here we just write down the final result:
\beq
J_{F} = exp{\; \frac{\pi}{2} \int d^2x \Phi_{c} \Box \Phi_{c}}
\label{17}
\eeq

\noindent We then get
\beq
{F_{2n}}^2(x_{1},...,x_{2n})= J_{F}(x_{1},x_{2},...,x_{2n}) 
\label{18}                                                                          
\eeq
Therefore we see that, in our formulation, the squared multipoint correlator can 
be identified with a fermionic Jacobian, exactly as in the 2-point case.\\
At this stage one has to solve the system of differential equations for $\Phi_{c}$ 
and $\eta_{c}$, obtained by replacing (\ref{14}) in (\ref{11}) and (\ref{12}). 
Finally, by inserting the result in (\ref{17}) and (\ref{18}), one obtains\\
\beq
F_{2n}(x_{1},...,x_{2n})=\left( \frac{\prod_{even}|x_{ij}|}
{\prod_{odd}|x_{ij}|} \right)^{\frac{1}{4}} 
\label{19}                                                                          
\eeq

\noindent where $i>j$ and even (odd) refers to a constraint on $i+j$; 
 $i, j = 1, 2, ... 2n$. 
We have also set an ultraviolet cutoff, which divides the coordinate 
differences, equal to $1$. This formula exactly coincides with the famous 
Kadanoff-Ceva's result \cite{KC}.\\

Let us now study the Baxter model \cite{Bax}, which can be considered as
two Ising systems interacting through their spin variables (this model is
related, through a duality transformation, to the Ashkin-Teller model 
\cite{AT}). As shown by Luther and Peschel \cite{LP}, the scaling limit of this 
model is described by the Thirring \cite{Thirring} interaction:\\
\beq
{\cal L}_{int} = - {\lambda}J_{\mu}J_{\mu} 
\label{20}                                                                          
\eeq
where, as before, $J_{\mu}$ is the Dirac fermionic current and the coupling 
constant $\lambda$ is proportional to the four-spin coupling of the original 
lattice model. The Baxter model is known to have two natural order parameters, the
magnetization and the polarization $<P> = <{\sigma_{i}s_{i}}>$, where
$\sigma_{i}$ and $s_{i}$ are the spin operators of each Ising system. 
In the continuous formulation the 2-point correlator for the polarization 
operator is given by\\
\[
<P(x)P(y)>_{\lambda} = <\sigma_{x}s_{x}\sigma_{y}s_{y}>_{\lambda}                                                                            
\]
where $< >_{\lambda}$ means v.e.v. with respect to the fermionic model defined 
by (\ref{20}). For $\lambda = 0$ the above expression becomes the squared Ising
correlator. This suggests the following identification:\\
\[
<P(0)P(R)>_{\lambda} = <exp\;{ \pi \int_{0}^{R} dz J_{0}(z)}>_{\lambda} 
\]
The r.h.s. of the precedent equation can be computed by employing a slightly
modified version of the method described above. Indeed, it is easy to show that
the introduction of an auxiliary vector field $A_{\mu}$ through a 
Hubbard-Stratonovich identity, allows to write\\
\beq
<P(0)P(R)>_{\lambda} = \frac{Z}{Z'} 
\label{23}
\eeq
with \\
\beq
Z = \int {\cal D}A_{\mu}\; e^{-\int d^{2}x {\frac{A^2}{2}}} \;
det \left(i \slp + (2\lambda)^{1/2}{\slB}\right)
\label{24}
\eeq
and
\beq
Z' = \int {\cal D}A_{\mu}\; e^{-\int d^{2}x {\frac{A^2}{2}}}\; 
det \left(i \slp + (2\lambda)^{1/2}{\slA}\right),
\label{25}
\eeq
where\\
\[
B_{\mu} = \epsilon_{\mu\nu}\partial_{\nu}{\Phi}_{B} + \partial_{\mu}{\eta}_{B}
\]                                                                          
\[
A_{\mu} = \epsilon_{\mu\nu}\partial_{\nu}\Phi + \partial_{\mu}\eta
\]                                                                          
\[
{\Phi}_{B} = \Phi + \frac{\pi}{\sqrt{2\lambda}}{\Phi}_{c}                                                                          
\]
\[
{\eta}_{B} = \eta + \frac{\pi}{\sqrt{2\lambda}}{\eta}_{c}.                                                                           
\]
Let us stress that, in contrast to the previous calculation of the Ising
correlator, in the present case one has to consider quantum fields $\Phi$ and 
$\eta$ whose dynamics plays a crucial role in the following computation. 
Concerning the classical functions $\Phi_{c}$ and $\eta_{c}$, they can be 
determined exactly as in the Ising case, i.e. using formulae (\ref{11}),
(\ref{12}) and (\ref{14}) for $n=1$.\\
We shall now turn to treat the fermionic determinants appearing in (\ref{24})
and (\ref{25}) by means of decoupling changes of fermionic variables, similar 
to the one defined by equations (\ref{15}) and (\ref{16}), but with parameters 
${\Phi}_{B}$ and ${\eta}_{B}$ in the form:
\[
\Psi = e^{-\sqrt{2\lambda}(\gamma_{5}\Phi_{B}+i\eta_{B})}\; \chi
\]
\[
\bP = {\bC}\; e^{-\sqrt{2\lambda}(\gamma_{5}\Phi_{B}-i\eta_{B})}
\]
The corresponding Jacobian is given by
\[
J_{F} = exp\;{ \frac{\lambda}{\pi} \int d^2x(\Phi+{\frac{\pi}{\sqrt{2\lambda}}} 
\Phi_{c}) \Box (\Phi+{\frac{\pi}{\sqrt{2\lambda}}} \Phi_{c})}
\]
Of course, this result must be used in (\ref{24}), whereas the same expression, 
but with $\Phi_{c}=0$ is to be employed in (\ref{25}). In so doing one readily 
discovers that, due to the fact that $J_{F}$ does not depend on
the field $\eta$, this field becomes decoupled from $\Phi$ in both $Z$ and
$Z'$. As the corresponding functional integrals over $\eta$ coincide, they
cancelled out when performing the quotient in equation (\ref{23}) and one
then gets

\[
<P(0)P(R)>_{\lambda} = <P(0)P(R)>_{0}\; <e^{\sqrt{2\lambda}\int d^2x{\Phi}
{\partial}_{\mu}{\partial}_{\mu}{\Phi}_{c}}> 
\]
where the first factor in the r.h.s. corresponds to the doubled
Ising correlator, whereas the second one  is a v.e.v. to be evaluated for a
model of free scalars $\Phi$ with Lagrangian density given by
\[
{\cal L} = (\frac{1}{2}+\frac{\lambda}{\pi})\partial_{\mu}\Phi\partial_{\mu}\Phi
\] 
As it is well-known this computation can be done by a standard shift in the
bosonic variable $\Phi$. The final result is \\  
\beq
<P(0)P(R)>_{\lambda} = (\frac{a}{R})^{2{\Delta}_{P}}
\label{34}
\eeq
where $a$ is an ultraviolet cutoff and $\Delta_{P}$ is the critical exponent
associated to the polarization operator, for which we get:\\
\beq
\Delta_{P} =\frac{1}{4} \frac{1}{1+\frac{2\lambda}{\pi}}\\
\label{35}
\eeq
Recalling the results for the energy-density ($\epsilon$) and the crossover 
($Cr$) operators \cite{Na} \cite{DK}, one obtains\\
\beq
4\Delta_{P} = \Delta_{\epsilon} =({\Delta_{Cr}})^{-1}\\
\label{36}
\eeq
which is the relation predicted by several authors \cite{E} \cite{K} and first
derived by Drugowich de Felicio and Koberle \cite{DK} in the operator 
framework.\\

In summary, we have extended a functional approach \cite{Na}, previously used to
compute 2-point functions in 2d critical systems, to the case in which 
multipoint correlators are considered. In particular, we provided an
alternative derivation of Kadanoff and Ceva's result \cite{KC} for the $2n$-spin
on-line function. Our contribution can be viewed as a complement to previous
works based on operational bosonisation, where 4-point functions were 
explicitly calculated \cite{ZI} \cite{Schroer}. We feel that our formulation
could be more practical when considering, for instance non-critical 
correlations \cite{Dotsenko}. Indeed, in this case one expects to have a
temperature-dependent ("massive") determinant, that can be easily handled
by following the perturbative strategy of ref.\cite{Naon}. The study of 
multipoint correlators in the defected Ising model \cite{defected} can be also 
envisaged in our scheme.\\
We have also computed the 2-point function describing the critical fluctuations
of the Baxter polarization operator. Thus we obtained its corresponding 
critical index. This completed the path-integral proof of the relationship
between energy-density, crossover and polarization exponents, which had been
initiated in ref.\cite{Na}.

\vspace{1cm}

\noindent {\bf Acknowledgement}\\
We are grateful to Fundaci\'on Antorchas (Argentina) for financial support.\\
We thank D.Cabra for a careful reading of the manuscript and useful suggestions.\\
One of us (CN) also thanks E.Fradkin for his explanation on the gauge symmetry
in the 2d Ising lattice.\\

\end{document}